\documentclass[preprint,12pt]{elsarticle}
\usepackage{amsmath,amssymb,siunitx,graphicx,hyperref,physics}
\usepackage{cleveref}
\usepackage{booktabs}
\usepackage{subcaption}
\usepackage{multirow}
\usepackage{pdflscape}
\biboptions{numbers,sort&compress}

\begin{document}
\setlength{\parindent}{0pt}  
\setlength{\parskip}{1em}    
\makeatletter
\def\ps@pprintTitle{%
   \let\@oddhead\@empty
   \let\@evenhead\@empty
   \let\@oddfoot\@empty
   \let\@evenfoot\@empty
}
\makeatother
\title{An analytical model for gold nanoparticle radiosensitisation}

\author[inst1,inst2]{Pedro Teles}\corref{cor1}\corref{cor2}\corref{cor3}

\affiliation[inst1]{organization={Departamento de Física e Astronomia da Faculdade de Ciências da Universidade do Porto, Porto, Portugal}}
\affiliation[inst2]{organization={Centro de Investigação do Instituto Português de Oncologia do Porto,Porto, Portugal}}

\cortext[cor1]{Corresponding author: ppteles@fc.up.pt}
\cortext[cor2]{A previous version of this paper included a derivation which integrated over negative doses. This was clearly an oversight and has now been corrected. Other typos and small corrections have also been performed.}
\cortext[cor3]{The preprint has now been further developed with some of the derivations more accurately explained and the supplementary material expanded.}

\begin{abstract}

In this paper, we derive a variance-driven Local‑Effect‑Model ($\sigma$-LEM) to predict  radiosensitization due to gold nanoparticles (AuNP). Assuming that the number of Au photo‑ionisations scales strictly with particle volume \(V_{\mathrm{NP}}\propto R^{3}\), a linear relation between dose enhancement ratio and concentration is achieved ($DER=1+K_c c$), in which $K_c$ is a beam quality and nucleus size-specific term, and $c$ is the concentration in mM.
Furthermore, assuming that the cascade energy deposition is log‑normally distributed, the enhanced dose in each target voxel can be written as \(D_{\text{enh}}=D_{0}\exp(\sigma Z)\) with \(Z\sim\mathcal N(0,1)\) and width \(\sigma=\sqrt{2\ln(1+Kc)}\).

Assuming a linear-quadratic (LQ) dose response, a relation between cell survival and dose can be derived. Despite no closed form for the log-normal distribution, averaging over the entire domain using first and second order moments leads to three possible closed forms: variance‑only, mixed‑term,  and  second-order. These three variants adapt well to low-concentration, mid-concentration, and high-concentration regimes.

The model was tested for Bovine aortic endothelial cells (BAEC) results taken from a Local Effect Model (LEM) and experimental values.The model agrees within $\le 2.5\%$ with the experimental and LEM data, but presents significant changes to the conceptual results obtained with the LEM. In particular, that AuNP dose enhancement is mostly $\alpha$-driven, as posited previously by other authors. These findings are further developed in the manuscript. 

The theoretical framework presented here  collapses radiobiological outcomes to three experimentally controllable variables — beam quality, nucleus size, and intracellular concentration $c$ — while retaining mechanistic fidelity.  Additional tests should be made to further confirm the validity of the model.
\end{abstract}

\maketitle

\section{Introduction}

Gold nanoparticles (AuNPs) (and other high-Z elements) have been the subject of extensive research for cancer treatment applications \cite{li2020, Li2020Corrigendum, Rabus2023, Thomas2024}. When irradiated, due to a heightened photoelectric cross-section coupled with high Auger electron yields,  AuNPs enhance the local dose in their surrounding, as produced Auger cascades  deposit orders‑of‑magnitude more energy within tens of nanometres than the original photon beam alone \cite{brown2017,chithrani2010gold,rahman2009}.  

The most effective model to describe AuNP radiosensitization is the Local Effect Model (LEM). LEM was originally developed for heavy-ions, with  McMahon \textit{et al.}\ first modifying it to score nanoscale radial dose profiles of gold nanoparticles (AuNPs), predicting surviving fractions that agreed with clonogenic data for 1.9 nm AuNPs under $105$–$220\,$kVp photons \cite{McMahon2011}.  Their analysis showed that the mean number of Au photo-ionisations scales with NP volume ($V_{\mathrm{NP}}\!\propto\!R^{3}$) but also that for equal mass many small NPs seem to have the same biological effect than a few large ones, showing that, besides the  AuNP sizes, concentration would play a significant role as well \cite{McMahon2011}. 

Lechtman \textit{et al.}\ combined full-track Monte Carlo (MC) energy deposition around 50 nm AuNPs with a voxelised linear–quadratic (LQ) survival calculation and reproduced prostate-cell SER within 1 \% \cite{lechtman2013}.   

Similar MC+LQ studies now exist for proton‐beam AuNPs \cite{Kirkby2017} and for kilovoltage gadolinium NPs (GdNPs) \cite{Wu2023}.  These models are very accurate, requiring the calculation of voxel doses using Monte Carlo  simulations around the nanoparticle.

In this work, we develop a completely theoretical $\sigma$-LEM,  derived exclusively from physical first principles, establishing a relationship between beam quality and stochastic local dose enhancement, showing that, at first-order, DER varies exclusively with concentration. By assuming a log-normal distribution for the dose enhancement, we average over the entire domain to obtain closed-form macroscopic equations for DERs and survival curves under different concentration regimes, showing that the first-order moment (with or without a mixed term) captures the lower to intermediate concentrations well, while the second-order moment aligns well with higher concentrations. 

The developed model was tested with previously published experimental and LEM results \cite{rahman2009, brown2017}  yielding good accuracy, although predicting an $\alpha$-driven enhancement, different from the prediction of the LEM.

Finally, the $\sigma$-LEM model has been developed for X-ray activation, but can potentially be adapted to other beams (protons, heavy-ions).

\section{Methodology}

The $\sigma$--LEM model was derived from physical first principles.
Spectrum-weighting, fitting, table and figure generation were all performed with the use of python scripts.  Experimental BAEC survival data were digitised with WebPlotDigitizer \cite{rohatgi2024webplotdigitizer}.  A single energy--independent constant, $\varepsilon_{\mathrm{cas}}N_{\mathrm{Auger}}\approx1\times10^{-5}\,\text{Gy}$, was obtained by minimising the Root Mean Square Error between model and synchrotron DERs.

Three analytic survival curves (variance--only, mixed, second--order) were derived by successive expansions in~$\sigma$ and Gaussian averaging. Truncation to avoid negative doses was necessary in the case of the first-order developments. The analytical curves were then cast into LQ form by weighted least squares, and compared.

All analytical steps were verified using ChatGPT-4o (OpenAI, 2025) and Claude 4.0 (Anthropic, 2025). After this verification, they were further validated using  Wolfram Cloud (Mathematica), and ultimately verified by the author. The author assumes full responsibility for the derivations. ChatGPT and Claude were also used as tools to enhance  Python scripts.

\sloppy Code and reproducibility notebooks are available at https://doi.org/10.5281/zenodo.15734277.

\section{Results}

\subsection{Baseline ionisations per Gray and primary photo-absorptions in one AuNP}

For a broad X‑ray field under charged‑particle equilibrium (CPE) the homogeneous dose deposited in tissue is
\begin{equation}
  D=\expval*{\Bigl(\tfrac{\mu_{\mathrm{en}}}{\rho}\Bigr)_{tissue}E}_{\Phi}\,\Phi,
  \label{eq:D0}
\end{equation}
\sloppy where \(E\) is photon energy, \(\Phi\) the incident fluence (photons~cm\(^{-2}\)), and \(\langle(\mu_{\mathrm{en}}/\rho)_{tissue}E\rangle\) is evaluated over the fluence spectrum.  Setting \(D=1\,\mathrm{Gy}\) defines the reference fluence
\begin{equation}
  \Phi_{1\,\mathrm{Gy}}=\frac{D_{1Gy}}{\expval*{\bigl(\mu_{\mathrm{en}}/\rho\bigr)_{tissue}E}_{\Phi}}.
  \label{eq:Phi1Gy}
\end{equation}

In a AuNP, the number of ionizations that occur in it, when traversed by an X-ray beam, is proportional to the number of photons that interact with the AuNP:

\begin{equation}
N_{ion} =  \frac{A_{AuNP}}{A_{source}}\cdot (1-e^{- <\mu_{phot}^{Au}>_{\Phi} \cdot  \bar{x}}),  
\label{eq:N_ion}
\end{equation}

where $A_{AuNP}$ is the area of the AuNP, $A_{source}$ is the area of the source used,  $<\mu_{phot}^{Au}>_{\Phi} $ is the mean photoelectric cross section of gold for the X-ray spectrum, and $\bar{x}$ is the mean chord length of the AuNP. Considering that $\bar{x}$ is extremely small, the previous equation is well approximated by the first term of the Taylor expansion, leading to \cite{Teles2025}:

\begin{equation}
N_{ion} =  \frac{A_{AuNP}}{A_{source}} \cdot <\mu_{phot}^{Au}>_{\Phi}  \cdot  \bar{x} + \mathcal{O}(\bar{x}^2),  
\label{eq:N_ion_II}
\end{equation}

Where the terms $\mathcal{O}(\bar{x}^2)$ are so small they can be neglected.

Considering that the mean chord length is \(\bar x = V_{\mathrm{AuNP}}/A_{\mathrm{AuNP}}\),  the number of ionizations per photon is given by:

\begin{equation}
N_{ion} =  \frac{A_{AuNP}}{A_{source}}\, <\mu_{phot}^{Au}>_{\Phi} \,  \bar{x} =  <\mu_{phot}^{Au}>_{\Phi}\frac{V_{AuNP}}{A_{source}},  
\label{eq:N_ion_II}
\end{equation}

meaning that the number of ionisations $N_{ion}$ is proportional to the volume of the AuNP, regardless of its shape, and also that $N_{ion} \propto R^3$.

If we multiply  equation \ref{eq:N_ion_II} by equation \ref{eq:Phi1Gy}, and by the total number of AuNPs $N_{AuNP}$, we obtain the number of ionizations per Gy:

\begin{align}
    N_{ion}^{Gy} &= D_{1Gy}\cdot\frac{<\mu_{\mathrm{phot}}^{\mathrm{Au}}>_{\Phi}}{\expval*{(\mu_{\mathrm{en}}/\rho)_{tissue}E}_{\Phi}}\cdot N_{AuNP} \cdot V_{AuNP}. \\
     &=\mathcal{F}(E)_{tissue}^{Au} \cdot N_{AuNP} \cdot V_{AuNP},
     \label{eq:N_1Gy_ion}
\end{align}

where $\mathcal{F}(E)_{tissue}^{Au}$ represents the number of ionizations per Gy per AuNP per volume of the AuNP. Using the typical units for dose $D(Gy)$, energy $E(MeV$ or $keV)$, $<\mu>\,(\mathrm{cm^{-1}})$, $<\frac{\mu}{\rho}>\,(\mathrm{cm^{2}/g})$, and the appropriate conversions ($MeV\,(\mathrm{or}\, keV)\rightarrow J$, and $g\rightarrow kg$), $\mathcal{F}(E)_{tissue}^{Au}$ has units of $\mathrm{cm^{-3}}$, easily convertible to  $\mathrm{nm^{-3}}$.

The number of ionisations per Gy is proportional to the volume of the nanoparticle. This has clear implications, like the shape of the NP being irrelevant as long as the volume is the same. Also, because \(V_{\mathrm{NP}}\propto R^{3}\), the mean cascade dose scales strictly with \(R^{3}\). At the same time, this equation tells us that the number of ionisations is proportional to the number of AuNPs.

\subsection{Log-normal distribution of dose deposition}

Experimental studies across many fields have repeatedly shown that the dose deposition by a photoelectron is log-normally distributed \cite{Prieskorn:2014,Sakurai2003,Soffitta2001,Miller2004} 

In fact, if we assume that the enhanced dose deposited in a scoring voxel per photon history can be described as \cite{Williamson1987}:
\begin{equation}
    D_{enh} = \frac{1}{m_{\text{vox}}}\sum_{i=1}^{N_{e}}
S\!\bigl(E_i\bigr)\,
\ell_i\
\end{equation}

in which $N_{e}$ is the number of electrons produced by the ionizations inside the voxel, $S(E_i) \equiv -\mathrm dE/\mathrm dx$ is the stopping power, $\ell_i$ the fraction of the electron’s track length that lies inside the voxel, dependent on stochastic mechanisms such as straggling and quasi-elastic interactions, are all positive values. Assuming that both are nearly independent, and considering the  geometric central-limit theorem, the product of these variables is log-normally distributed. In this picture, the voxel must be such that it is smaller than the range of Auger electrons.

The total number of electrons per Gy of baseline dose is $N_e=N_{ion}^{Gy}\cdot N_{Auger}$, where $N_{Auger}$ is the number of Auger electrons produced by an ionization event in the AuNP.

\subsection{Stochastic form and relation to macroscopic dose}

If local dose in the vicinity of the AuNP is log-normally distributed, it follows that
\begin{equation}
  D_{\text{enh}} = D_{0}\exp(\sigma Z),\qquad Z\sim\mathcal N(0,1),\label{eq:Dloc}
\end{equation}

meaning that for the distribution, 

\begin{equation}
  \ln \left(\frac{D_{\text{enh}}}{D_0} \right) = \sigma Z,\label{eq:Dloc_ln}
\end{equation}

 $\sigma$ is its standard deviation.

$D_0$ is the macroscopic dose so that the total number of ionizations for that dose is given by:

\begin{equation}
    N_{ion}^{total} = D_0\, N_{ion}^{Gy}.
    \label{eq:N_ion_total}
\end{equation}

\subsection{Absorbed dose to nucleus}

Now, considering that each ionisation that launches an Auger cascade  deposits \textit{on average }\(<\varepsilon_{\text{cas}}N_{e}>\,\mathrm{Gy}\) per Gy of baseline, then
\begin{equation}
  D^{(1)}_{\text{extra}} = <\varepsilon_{\text{cas}} N_{e}>,
  \label{eq:D_extra}
\end{equation}

is the extra dose per baseline Gy due to the ionizations in one AuNP.

Then defining $D_{\text{extra}}$ as the mean additional (macroscopic) dose due all AuNPs, and using equations \ref{eq:N_1Gy_ion} and\ref{eq:N_ion_total}, we write this as:

\begin{equation}
  D_{\text{extra}} = D_0\,<\varepsilon_{\text{cas}}\,N_{Auger}>\,\mathcal{F}(E)_{tissue}^{Au}\,V_{AuNP}\,N_{AuNP}.
  \label{eq:D_extra_II}
\end{equation}

where 

\begin{equation}
    <\varepsilon_{\text{cas}}\,N_{Auger}>=\int_{R_p}^{\infty}\Theta\!\bigl(R_{\text{eff}}-x\bigr)\,\int_0^{D_{max}}D^{(1)}(E,x)\Phi'^{(1)}(E,x)\,dE\,dx,
    \label{eq:epsilon_cas}
\end{equation}

is the total enhanced dose due to the Auger cascade enhancement per baseline Gy. 

The term $D^{(1)}(E,x)$ is the dose kernel normalized per fluence per energy, which can typically be obtained by Monte Carlo track-structure analysis and $\Phi'^{(1)}(E,x)$ is the fluence differential per energy (E) and distance to the AuNP's centre (x). 

The absorbed dose per particle fluence is plotted in figure 11b of \cite{Thomas2024}, showing a steep decrease up to about $R_{\text{eff}}\sim 500 nm$. This led the authors to conclude that nearest neighbour AuNPs will not be affected by the localised extra-dose produced by each of them, thus making the local AuNP effect confined to a radius $R_{\text{eff}}$ around the AuNP even for very high concentrations, which we simplify by introducing the Heaviside function $\Theta\!\bigl(R_{\text{eff}}-x\bigr)$. One immediate implication of this is that these single extra doses are additive, and no real synergistic effect exists between them. This assumption is also present in this derivation, where the term $D_{extra}$ in eq \ref{eq:D_extra_II} is calculated as the total enhanced dose due to the entirety of the AuNPs present in the cytoplasm.

The extra dose to the nucleus is a fraction of $D_{extra}$:

\begin{equation}
    D_{nuc,extra}=D'_{extra}.
\end{equation}

This  can be estimated by assuming that the AuNPs are uniformly and randomly distributed in a spherical shell around the nucleus such that, if $r$ is the distance to the nucleus centre:

\begin{equation}
    R_{nuc} < r < R_{cell}.
    \label{eq:r_distance}
\end{equation}

From this, we can state from that only those AuNPs whose $r$ lies within a spherical shell ($R_{\text{eff}}-R_{nuc}\sim 500$ nm) will deposit dose in the nucleus:

\begin{equation}
    R_{nuc} < r < R_{eff}.
    \label{eq:r_eff}
\end{equation}

The fraction $f$ of the total AuNPs that lie within that shell is simply the ratio between the two shells:

\begin{equation}
    f_{hit} = \frac{\frac{4}{3}\pi\left((R_{nuc}+R_{eff})^3-(R_{nuc}+R)^3\right)}{\frac{4}{3}\pi\left(R_{cell}-R_{nuc}\right)^3} = \frac{(R_{nuc}+R_{eff})^3-(R_{nuc}+R)^3}{\left(R_{cell}-R_{nuc}\right)^3}
\end{equation}

Finally, the dose from the shell of $f_{hit}$ AuNPs that actually falls inside the nucleus is given by the two-sphere intersection volume, in which $(R_{\text{nuc}} - R_{\text{eff}}) \leq r \leq R_{\text{nuc}} + R_{\text{eff}}$:

\begin{equation}   
V_{\cap}(r) = \frac{1}{12r} \pi \left( R_{\text{nuc}} + R_{\text{eff}} - r \right)^2 \left[ r^2 + 2r(R_{\text{nuc}} + R_{\text{eff}}) - 3(R_{\text{eff}} - R_{\text{nuc}})^2 \right],
\end{equation}

where the fraction will be given by:

\begin{equation}
    \kappa(R) =   \frac{ \displaystyle\int_{R_{\text{nuc}} + R}^{R_{\text{nuc}} + R_{\text{eff}}}  V_{\cap}(r)\, 4\pi r^2 \, dr }{ \displaystyle V_{\text{eff}}\int_{R_{\text{nuc}} + R}^{R_{\text{nuc}} + R_{\text{eff}}} \ 4\pi r^2 \, dr },
    \label{eq:Integral_R}
\end{equation}

In which $V_{\text{eff}}=\frac{4}{3}\pi R_{\text{eff}}^3$.

The solution to the integral in equation \ref{eq:Integral_R} has an analytical closed form which is shown in the supplementary material.

\begin{figure}[h!]
    \centering
    \includegraphics[width=0.8\linewidth]{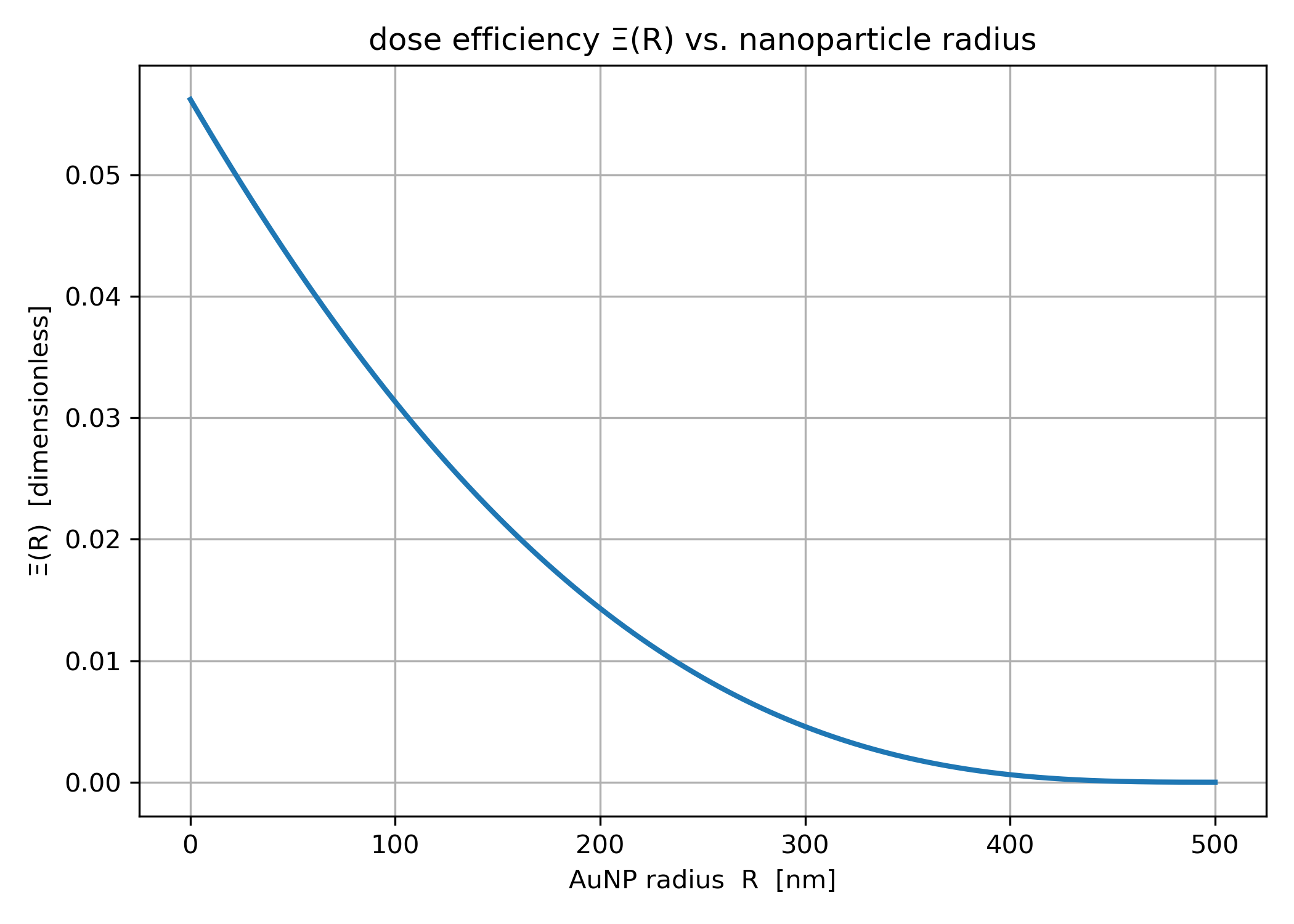}
    \caption{Dose efficiency to the nucleus as a function of nanoparticle radius}
    \label{fig:dose_efficiency}
\end{figure}

So the dose to the nucleus is given by:

\begin{equation}
    D_{\text{nuc,extra}}=\Xi(R)\,D_{\text{extra}},
\end{equation}

where the "dose efficiency" to the nucleus, $\Xi(R)=f_{hit}\,\kappa(R)$, is an R-dependent function. In figure \ref{fig:dose_efficiency} the value of $\Xi(R)$, for $R_{nuc}=5$ $\mu$m and $R_{cell}=10$ $\mu$m is shown, as a function of particle radius $R$. This function is monotonically decreasing in R, with a value of and tends asymptotically to 0 at $R_{eff}$. This is because of the assumptions made in the model. Dose in the nucleus is only possible if $r$ lies between $R_{nuc}+R<r<R_{nuc}+R_{eff}$. If $R=R_{eff}$ both $f_{hit}(R=0)$ and $\kappa(R=0)$ are zero, both because we are assuming that the Auger enhancement has a hard limit at $R_{eff}$, hence if $R=R_{eff}$ this enhancement is completely absorbed by the AuNP itself, and because a particle that is too large cannot approach the nucleus close enough because its own centre is too far out in relation to $R_{eff}$ for the particle's dose enhancement to ever reach the nucleus. For realistically sized AuNPs, the value of $\Xi(R)$ varies only just slightly, from circa 3\% for 100 nm to about 5\% for 10 nm AuNPs.

Finally, plugging in equation \ref{eq:D_extra_II}, $D_{\text{nuc,extra}}$ can be written as

\begin{equation}
    D_{\text{nuc,extra}}=D_0\,\Xi(R)\,<\varepsilon_{\text{cas}}\,N_{Auger}>\,\mathcal{F}(E)_{tissue}^{Au}\,V_{AuNP}\,N_{AuNP}
\end{equation}

\subsection{Dose enhancement ratio}

\begin{equation}
    <D_{enh}>=D_0+D_{\text{nuc,extra}}.
\end{equation}

Then, the dose enhancement ratio can be written as:

\begin{equation}
    \mathrm{DER}=\frac{<D_{enh}>}{D_0}=1+\frac{D_{\text{nuc,extra}}}{D_{0}} = 1+KN_{AuNP}R^{3},
    \label{eq:DER}
\end{equation}

Where all the constant terms are flushed into:

\begin{equation}
    K(E)=\frac{4\pi}{3}\,\Xi(R)\,<\varepsilon_{\text{cas}}\,N_{Auger}>\,\mathcal{F}(E)_{tissue}^{Au}\,,
    \label{eq:K_E}
\end{equation}

where $\mathcal{F}(E)_{tissue}^{Au}$ has already been defined in equation \ref{eq:N_1Gy_ion}, $K(E)$ is a function of energy and comes in units of $\mathrm{NP^{-1}\,nm^{-3}}$. 

If one wants to specify the intracellular concentration $c\,[\mathrm{mM}]$, knowing that the NP count in one cell is
\begin{equation}
  N_{\text{AuNP}}=c\,V_{\text{cyt}}\,\frac{M_{\text{Au}}}{\rho_{\text{Au}}}\,\frac{3}{4\pi}\,\frac{1}{R^{3}}\,.
  \label{eq:NPcount}
\end{equation}

where $M_{\text{Au}}$ is the total mass of gold, $\rho_{\text{Au}}$ is the density of gold, and $V_{\text{cyt}}$ is the cell nucleus cytoplasm.

Replacing equation \ref{eq:NPcount} into equation \eqref{eq:DER} eliminates the $R^{3}$ term and defines a concentration prefactor $K_c$:

\begin{equation}
  K_{c}(R,E)=\Xi(R)<\varepsilon_{\text{cas}}\,N_{\text{Auger}}>\,\mathcal{F}(E)_{tissue}^{Au}\,V_{\text{cyt}}\,\frac{M_{\text{Au}}}{\rho_{\text{Au}}}\qquad[\mathrm{mM^{-1}}]\,.
  \label{eq:KcE}
\end{equation}
Hence
\begin{equation}
 \mathrm{DER}=1+K_{c}(R,E)\,c\,.
  \label{eq:DERc}
\end{equation}

This is an interesting finding as it corroborates previous findings that reported dose enhancement was mostly due to concentration. Some authors reported that smaller AuNPs in higher concentrations closer to the nucleus had more enhancement than larger ones \cite{McMahon2011}.

In fact, recent research seems to take this a step further by starting off their derivations under the fundamental assumption that a similar equation to equation \ref{eq:DERc} (linear with concentration) holds true at a fundamental level in AuNP enhancement \cite{Blind2025}. This finding was used by the authors to enhance Monte-Carlo simulations.

the expression for $K_c(R,E)$ reveals that this constant will be dependent on the cell dimensional properties ($V_{cell}$, $V_{nuc}$ and $V_{cyt}$), and AuNP radii due to the term $\Xi(R)$, which for constant cell and effective radii gives a variation with AuNP size, and beam quality, in the term $\mathcal{F}(E)_{tissue}^{Au}$, as explained previously.

Finally, knowing that, for a standard normal variable  \( Z \),
$\mathbb{E}\left[e^{\sigma Z}\right] = e^{\sigma^2 / 2}$ \cite{abramowitz1965handbook}, then

\begin{equation}
    \langle D_{\text{enh}} \rangle = D_0\, e^{\sigma^2 / 2}.
    \label{eq:mean_Denh}
\end{equation}

. 

Solving for $\sigma$, and using equation \ref{eq:DER}:

\begin{align}
    \sigma^2 &= 2\ln \left( \frac{<D_{enh}>}{D_0} \right) = 2\ln (\text{DER})=2\ln (1\,+\,K_cc) \\
    \sigma &=\sqrt{2\ln (1\,+\,K_cc)}
    \label{eq:sigma}
\end{align}

Equation \ref{eq:sigma} shows that the stochastic width
$\sigma$ is now a deterministic function of two measurable
quantities: beam-specific, and cell specific $K_c$ and intracellular molar concentration $c$. The 
particle radius $R$ dependence disappears, which is consistent with results showing that higher concentrations of smaller AuNPs yield higher DERs than larger AuNPs \cite{McMahon2011}.  No voxel histograms or Monte-Carlo dose maps are required once $K_c$ has been fixed for the spectrum.

To note that a more rigorous treatment would parametrise the enhanced dose as $D_{\text{enh}} = D_0 \exp(\mu + \sigma Z)$ with a separate ($\mu$) term. 

For this more general form, the expectation value would be $\mathbb{E}[D_{\text{enh}}] = D_0 \exp(\mu + \sigma^2/2)=D_0'\exp(\sigma^2/2)$, with $D_0'=D_0\exp(\mu)$ corresponding to a deterministic dose shift. Solving this now leads to one single equation for two unknown variables:

\begin{equation}
    \mu+\frac{\sigma^2}{2}=\ln(1+K_cc)
    \label{eq:two_unknowns}
\end{equation}

To solve this problem, we can define the right side of equation \ref{eq:two_unknowns} as the macroscopic gain $L$, such that:

\begin{equation}
    L=\ln(1+K_cc),
    \label{eq:mac_gain}
\end{equation}

and $\mu=L-\frac{\sigma^2}{2}$. Introducing a linear factor $k$ which quantifies what fraction of the gain is variance-related:

\begin{equation}
    \sigma^2= k\,L.
    \label{eq:sigma_2}
\end{equation}

$k$ functions as a "splitting factor", quantifying what fraction of the total gain is stochastic or deterministic:

\begin{equation}
    \frac{\sigma^2}{\mu} = \frac{k}{1-k/2},
    \label{eq:splitting_factor}
\end{equation}

which leads to a scaling of the form $\sigma^2 = k \ln(1 + K_c c)$, and $\mu=(1-\frac{k}{2})\ln(1 + K_c c)$. 

Values of $k>2$ would lead to the median $D_0\exp(\mu)<D_0$, a physically implausible scenario which can therefore be discarded, setting boundaries for $0\le k \le2$.

A completely deterministic dose shift would be attained with $k=0$, leading to $\sigma=0$. The dose enhancement would strictly be due to the mean $\mu=\ln(1+K_cc)$, contradicting the evidence of discrete dose depositions around the AuNP. For $k=1$, the enhancement would split evenly between the deterministic shift and stochastic enhancement. Finally, for $k=2$, $\mu=0$, and the fully stochastic regime modelled by equation \ref{eq:sigma} is retrieved.

To note that setting $\mu=0$ does not imply the mean dose stays at $D_0$, given that for a log-normal the mean is given by equation \ref{eq:mean_Denh}, so variance ($\sigma^2$) alone can raise the mean. 

The value of $k=2$ maintains consistency with our physically motivated scenario, in which dose enhancement arises strictly from increased stochastic fluctuations, based on the well-known Monte Carlo dose distribution studies around AuNPs.

The previously obtained expressions for $K_c(E)$, DER and $\sigma$ can be directly compared with the values obtained for a monoenergetic photon beam irradiating Bovine aortic endothelial cells (BAEC) at different energies (30, 40, 50, 60, 70, 81, and 100 keV) at a concentration of AuNPs of $c=1$ mM \cite{rahman2014optimal}. The values used to estimate $K_c$ were  $\rho_{Au} = 19.3\,\mathrm{g\,cm^{-3}}$
$M_{Au} = 196.97\,\mathrm{g\,mol^{-1}}$,$V_m = \tfrac{M_{Au}}{\rho_{Au}}\approx 10.2\,\mathrm{cm^{3}\,mol^{-1}})$, and $V_{cyt} = 5.0\times10^{-11}\,\mathrm{cm^{3}}$, whereas the photon cross-sections for gold and tissue were taken from the National Institute of Standards and Technology, USA (NIST)\cite{nist_xray_coefficients}.

To note that the authors used the dose enhancement factor $DEF_{80\%}$ at 80\% which is calculated through the equation:

\begin{equation*}
    DEF_{80\%}=\frac{\mathrm{Dose\,at\,survival\,fraction \,of\, 80\%\,with\,no\,AuNPs}}{\mathrm{Dose\,at\,survival\,fraction \,of\, 80\%\,with\,AuNPs}}=\frac{D_0^{80}}{D_{Au}^{80}}
\end{equation*}

which can be related to $DER=\frac{<D_{enh}>}{D_0}$ as defined in equation \ref{eq:DER} by equating the survival fractions using the LQ model:

\begin{equation*}
    S_{80}=\exp\left(-\alpha D_0^{80}-\beta (D_0^{80})^2\right)
\end{equation*}

and

\begin{equation*}
    S_{80}=\exp\left(-\alpha D_{Au}^{80}-\beta (D_{Au}^{80})^2\right)
\end{equation*}

Knowing that for any survival fraction, but in particular at 80\%, $D_{Au}^{80}=DER\cdot D_0^{80}$, it is easy to see that:

\begin{equation*}
    DER=DEF
\end{equation*}

Therefore relating the radiobiological  $DEF$ with the physical $DER$ concept of equation \ref{eq:DER}.

\begin{table}[h!]
    \centering
        \begin{tabular}{cccc}
        \hline
        Energy (keV) & DER$_\text{exp}$ (DEF$_{80\%}$) & DER$_\text{model}$ & \% Diff \\
        \hline
        30 & 1.74 & 4.04 & +132.2\% \\
        40 & 3.47 & 3.11 & -10.4\% \\
        50 & 2.77 & 2.24 & -19.1\% \\
        60 & 3.06 & 1.51 & -50.7\% \\
        70 & 1.30 & 0.96 & -26.2\% \\
        81 & 2.58 & 3.02 & +17.1\% \\
        100 & 1.35 & 1.44 & +6.7\% \\
        \hline
        \end{tabular}
    \caption{DER values obtained using this work's model and the synchrotron results of \cite{rahman2014optimal}}
    \label{tab:DER_comparison}
\end{table}

\begin{figure}[h!]
    \centering
    \includegraphics[width=0.99\linewidth]{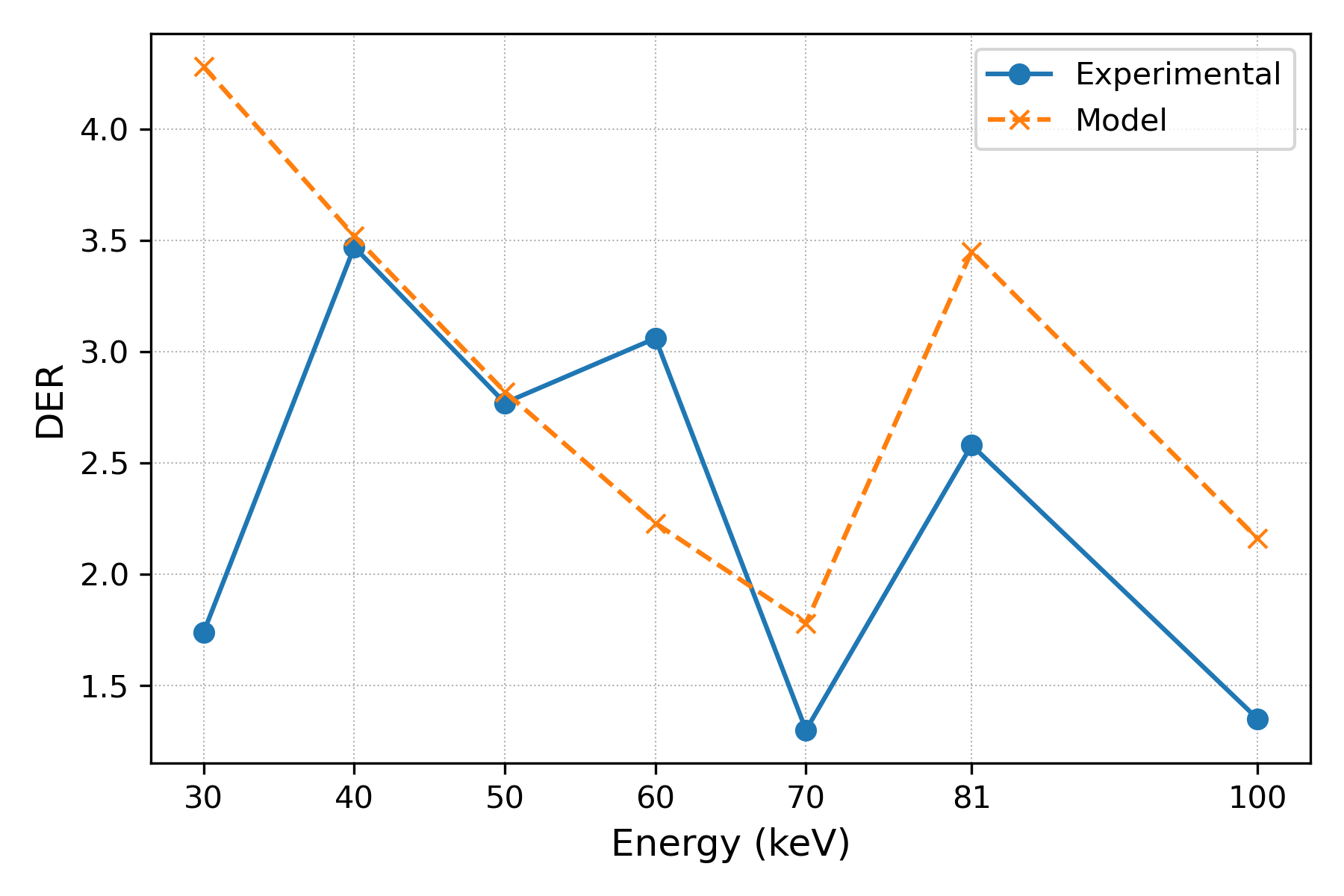}
    \caption{Graphical comparison of model DER vs syncrhotron experimental DER}
    \label{fig:DER_comparison}
\end{figure}

The comparison is presented in table \ref{tab:DER_comparison}, together with a graphical comparison shown in figure \ref{fig:DER_comparison}, where $DER_{\text{exp}}$ equates to the experimental $DEF_{80\%}$ given in the paper.

Despite its minimal structure, the present cascade-based model reproduces the published DERs to better than a factor $\approx$2 (RMSE $\approx$ 0.9), with a clear prediction of the dip at 70 keV. However, there’s some discrepancy in the lower energy DER comparison, perhaps because low-energy measurements ($\lesssim$ 40 keV) are more sensitive to experimental uncertainties.

The two sole free quantities are the product $<\varepsilon_{cas}N_{Auger}>$, i.e. the mean lethal dose delivered to nuclear DNA per gold photo-ionisation, $Xi(R)=0.05594$, which for this case ($R$=0.95 nm) and using the same radii as above..
Direct experimental or in-silico estimates for this quantity do not exist; literature values for both 
$\epsilon_{\mathrm{cas}}\ (10^{-7}\text{--}10^{-4}\,\mathrm{Gy\,cascade}^{-1})$
and 
$N_{\mathrm{Auger}}\ (10\text{--}30\ \text{electrons})$
span orders of magnitude. Calibrating the product to the synchrotron dataset yields $<\epsilon_{\mathrm{cas}}\, N_{Auger}>\approx 2.2\times10^{-4}\,\mathrm{Gy}$, well within the plausible range. Further calibration could entail using a more accurate value for a BAEC nucleus.

\begin{figure}[h!]
    \centering
    \includegraphics[width=0.99\linewidth]{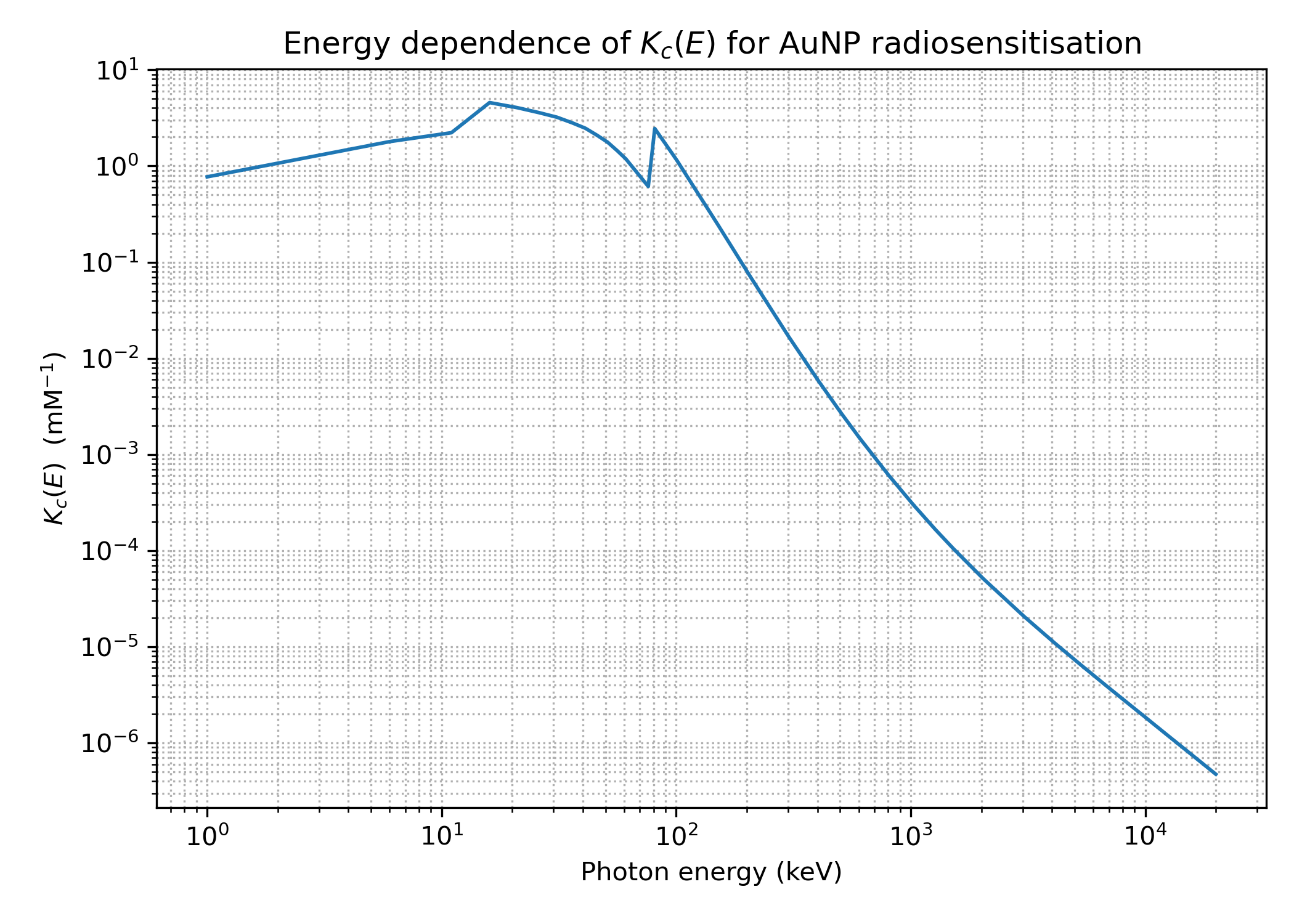}
    \caption{$K_c(E)$ as a function of energy (keV) in log-log scale}
    \label{fig:K_c_vs_en}
\end{figure}

In figure \ref{fig:K_c_vs_en}, $K_c(R,E)$ is provided as a function of energy (1keV-20MeV), assuming monoenergetic beam values, and an $R=0.95$ nm, just as in \cite{rahman2014optimal}.

Taken together, the agreement supports the view that a single, energy-independent calibration of $<\varepsilon_{cas} N_{Auger}>$, combined with known cross-sections and a realistic cellular dimensionality, is sufficient to predict DER within experimental scatter.

However, clinical beams are typically not monochromatic, and therefore the bracketed terms should be spectrum-weighted. 

We determined the value of $K_c(R,E)$ for two kilovoltage (kV) spectra - 50 and 100 kVp, determined with SpekPy \cite{SpekPy}, using an anode angle of $20^\circ$ and a filtration setup of 0.8 mm of beryllium (Be) + 3.9 mm of aluminum (Al), and for four mega-voltage (MV) spectra as reported by \cite{sheikhbagheri2002beam}.

The results are summarized in table \ref{tab:kc_values}.

\begin{table}[h!]
\centering
\caption{Spectrum-averaged radiosensitisation constant
         \(K_{\rm c}\) for the clinical photon beams used in this work.}
\begin{tabular}{@{}l c@{}}
\toprule
\textbf{Beam} & \(\boldsymbol{K_{\rm c}}\) \textbf{(mM$^{-1}$)} \\[2pt]
\midrule
\multicolumn{2}{@{}l}{\textit{kV beams}}\\
\hspace{1em}50 kVp  & \(3.22\)\\
\hspace{1em}100 kVp & \(2.11\)\\[6pt]
\multicolumn{2}{@{}l}{\textit{MV beams}}\\
\hspace{1em}Varian 10 MV   & \(1.23\times10^{-4}\)\\
\hspace{1em}Varian 6 MV    & \(3.48\times10^{-4}\)\\
\hspace{1em}Siemens 6 MV   & \(3.72\times10^{-4}\)\\
\hspace{1em}Elekta 6 MV    & \(3.01\times10^{-4}\)\\
\hspace{1em}Varian 4 MV    & \(2.24\times10^{-4}\)\\
\bottomrule
\label{tab:kc_values}
\end{tabular}
\end{table}

McMahon \textit{et al.}\ \cite{McMahon2011} noted that radiosensitisation rises with both
total Au mass (concentration) and AuNP volume, in such a way that the $R^3$ dependence is mitigated if smaller particles in higher quantities are used, which is observed in our model, given that because the $R^3$ terms cancel out in equation \ref{eq:DERc}, it only depends on the concentration.  In table \ref{tab:der_extended} values of DER are given for the spectra for three different concentrations. 

It can be seen that for kV beams the DER will be measurable. However, for mega-voltage beams, even with the lower-energy tail, the projected DERs would be negligible according to this model. This seems to contradict several experiments which have shown that DERs are not negligible in MV beams. This sets perhaps a limit of validity to the current model.

\begin{table}[h!]
\centering
\caption{DER predicted by
         $\mathrm{DER}=1+K_{\rm c}\,c$
         for three concentrations in mM}
\begin{tabular}{@{}l ccc @{}}
\toprule

 & 0.1 & 0.5 & 1.0 \\
\midrule
50 kVp     & 1.322 & 2.611 & 4.222  \\
100 kVp   & 1.211 & 2.055 & 3.111  \\[4pt]
Varian 10 MV        & 1.0000123 & 1.0000615 & 1.000123  \\
Varian 6  MV        & 1.0000348 & 1.000174 & 1.000348 \\
Siemens 6 MV        & 1.0000372 & 1.000186 & 1.000372  \\
Elekta  6 MV        & 1.0000301 & 1.000150 & 1.000301  \\
Varian 4  MV        & 1.0000224 & 1.000112 & 1.000224 \\
\bottomrule
\label{tab:der_extended}
\end{tabular}

\vspace{4pt}
\small
\end{table}

\subsection{Analytical survival curves}

Expanding equation \ref{eq:Dloc} to first order gives \cite{Teles2025}:

\begin{equation}
  D_{\text{enh}} = D_{0}(1+\sigma Z) + \mathcal{O}(\sigma^2),
  \label{eq:D_enh_first_order}
\end{equation}

in which we're going to neglect second order terms, the end result is in fact a gaussian distribution.

Now we're going to use the linear quadratic model to determine the survival curve for each $D_{enh}$:
\begin{equation}
    S_{enh} = e^{-\alpha D_{enh} - \beta D_{enh}^2}.
    \label{eq:S_fraction_single}
\end{equation}

Replacing $D_{enh}$ with equation \ref{eq:D_enh_first_order} yields for each term in equation \ref{eq:S_fraction_single} \cite{Teles2025}:

\begin{equation}
  \begin{aligned}
-\alpha D_{\text{enh}}
&= -\alpha D_{0}\bigl(1+\sigma Z\bigr)
   = -\alpha D_{0} \;-\; \alpha\sigma D_{0} Z, \\[4pt]
-\beta D_{\text{enh}}^{2}
&= -\beta D_{0}^{2}\bigl(1+2\sigma Z+\sigma^{2}Z^{2}\bigr)  \\[2pt]
&= -\beta D_{0}^{2} \;-\; 2\beta\sigma D_{0}^{2} Z
   \;-\;\beta\sigma^{2}D_{0}^{2}Z^{2}.
\end{aligned}
\label{eq:coefficients}
\end{equation}

The cross-term $2\sigma Z$ corresponds to a simultaneous energy deposit of both baseline photons and enhanced Auger electrons at the same voxel \cite{brown2017}, considered of low probability; therefore, it is often neglected \cite{Lechtman2017}.

Neglecting these terms leads to:

\begin{equation}
    S_{enh} \approx
    e^{
      -\alpha D_{0}(1+\sigma Z)\;-\;\beta D_{0}^{2}
      \;-\;\beta\sigma^{2}D_{0}^{2}Z^{2}
    },
    \label{S_fraction_III}
\end{equation}

The macroscopic survival fraction can be obtained by averaging over the Gaussian fluctuations.
\begin{align}
  S'_{\sigma}=\bigl\langle S_{enh} \bigr\rangle
  &= e^{-\alpha D_{0}-\beta D_{0}^{2}}\,
     \mathbb E_{Z}\!\bigl[
        e^{-\alpha\sigma D_{0}Z-\beta\sigma^{2}D_{0}^{2}Z^{2}}
     \bigr]                                                     \nonumber\\[4pt]
  &= e^{-\alpha D_{0}-\beta D_{0}^{2}}\,
     \frac{
           \exp\Bigl(
             \dfrac{\alpha^{2}\sigma^{2}D_{0}^{2}}
                   {2\bigl(1+2\beta\sigma^{2}D_{0}^{2}\bigr)}
           \Bigr)}
          {\sqrt{1+2\beta\sigma^{2}D_{0}^{2}}},
  \label{eq:Ssigma_D0}
\end{align}

where the Gaussian identity
$\mathbb E[e^{tZ+bZ^{2}}]=\frac{e^{\frac{t^{2}}{2(1-2b)}}}{\sqrt{1-2b}}$ for $b<\frac12$ \cite{abramowitz1965handbook}(eq. 7.4.32) was used (derivation given in Supplementary Material).

Also, given that  
$b=-\beta\sigma^{2}D^{2}$ is always negative, since 
$\beta,\sigma^{2},D^{2}>0$, then  $1-2b \geq 1$, making the hard limit of validity only happen if $2\beta\sigma^{2}D^{2} \rightarrow 1$, or $\beta \approx \frac{1}{2\sigma^{2}D^{2}}$, meaning an extremely low value of $\beta$.

Finally, in the equation, $\sigma=\sqrt{2ln(1+K_cc}$, $D_0$ is the baseline dose with no AuNPs, and $\alpha$,$\beta$ are the linear and quadratic coefficients of the LQ model.

This equation is completely defined for any given value of $K_c$ (beam  dependent), and the molar concentration $c$.

Another aspect to take into account is that the integration cannot include negative doses, so equation \ref{eq:Ssigma_D0} is in fact truncated at $z=-\frac{1}{\sigma}$, which introduces a correction factor $\mathcal R_{\rm trunc}$ to the equation, which is simply derived from truncating the derivation (see Supplementary Material): 

\begin{equation}
    \mathcal R_{\rm trunc}(\alpha,\beta,\sigma)
        \;=\;
\frac{\displaystyle
  \operatorname{erfc}\!\Bigl(
     \frac{\sqrt{1+2\beta\sigma^{2}D_{0}^{2}}}{\sqrt2}\,
     \Bigl[-\frac1\sigma+\frac{\sigma D_{0}\alpha}
                         {1+2\beta\sigma^{2}D_{0}^{2}}\Bigr]\Bigr)}
{\displaystyle
  \operatorname{erfc}\!\bigl(-\tfrac1{\sigma\sqrt2}\bigr)}
  \label{eq:Corr_factor}
\end{equation}

So that the equation for the survival curve is given by:

\begin{equation}
      S_{\sigma}= e^{-\alpha D_{0}-\beta D_{0}^{2}}\,
     \frac{
           \exp\Bigl(
             \dfrac{\alpha^{2}\sigma^{2}D_{0}^{2}}
                   {2\bigl(1+2\beta\sigma^{2}D_{0}^{2}\bigr)}
           \Bigr)}
          {\sqrt{1+2\beta\sigma^{2}D_{0}^{2}}}\mathcal\cdot R_{\rm trunc}(\alpha,\beta,\sigma)
        \label{eq:S_sigma_complete}
\end{equation}

The next step was to test the validity of the model. For that, we used the data provided by \cite{brown2017} as a source, with experimental data taken from \cite{rahman2009}. These two works were performed on BAECs using different X-ray beam qualities (Brown's work was used to model Rahman's results). Experimental values were extracted by making use of webplotdigitizer \cite{rohatgi2024webplotdigitizer} which allowed us to retrieve the experimental data points to within two significant figures. 

We compared our model against their results, obtained for the 100 kVp beam, for concentrations of 0.00, 0.25, 0.50, and 1.00 mM of AuNPs. The survival curve for our model was determined using equation \ref{eq:S_sigma_complete}, and the pre-calculated $K_c=2.11$ mM$^{-1}$ value for our own 100 kVp spectrum,  applied for the same concentration values. The LQ baseline coefficients were taken from \cite{brown2017} to be $\alpha=2.52\times10^{-2}\,\mathrm{Gy^{-1}}$ and $\beta=1.30times10^{-2}\,\mathrm{Gy^{-2}}$ for 1 mM. 

The application of equation \ref{eq:S_sigma_complete} yielded Dose Enhancement Factors at 80\% survival ($\mathrm{DEF_{80\%}}$ of 1.30, 1.51 and 1.74, with a percent deviation of +11.2\%, +2.3\%, and -10.4\% to the reported values \cite{rahman2009, brown2017} of 1.17, 1.47 and 1.94. The model thus seemed well calibrated to the intermediate concentration value (0.50 mM), but failed for the other two concentrations (0.25 and 1.00 mM). 

With these results, we decided to reintroduce the term $2\sigma Z$, meaning that we're now considering a non-negligible  probability of double baseline+enhance dose deposition on the same site. If one collects the terms of same order of $Z$ from equation \ref{eq:coefficients}, we can see that equation \ref{S_fraction_III} can be rewritten as:

\begin{equation}
    \begin{aligned}
        S_{enh} = &
        e^{
        -\alpha D_{0}-\alpha \sigma D_0 Z-2\beta\sigma D_0^2Z-\beta D_{0}^{2}
        -\beta\sigma^{2}D_{0}^{2}Z^{2}} \\
        = & e^{-\alpha D_0-\beta D_{0}^{2}}e^{-(\alpha - 2\beta D_0)\sigma D_0Z-\beta D_{0}^{2}},        
    \end{aligned}
    \label{S_fraction_mixed}
\end{equation}

If we make  $\alpha^{*}=\alpha + 2\beta D_0$, the equation becomes formally similar to \ref{eq:S_sigma_complete},  leading to:

\begin{align}
  S_{\sigma}^{\text{mixed}}=\bigl\langle S_{enh} \bigr\rangle
  &= e^{-\alpha D_{0}-\beta D_{0}^{2}}\,
     \mathbb E_{Z}\!\bigl[
        e^{-\alpha^{*}\sigma D_{0}Z-\beta\sigma^{2}D_{0}^{2}Z^{2}}
     \bigr]                                                     \nonumber\\[4pt]
  &= e^{-\alpha D_{0}-\beta D_{0}^{2}}\,
     \frac{
           \exp\Bigl(
             \dfrac{\alpha^{*2}\sigma^{2}D_{0}^{2}}
                   {2\bigl(1+2\beta\sigma^{2}D_{0}^{2}\bigr)}
           \Bigr)}
          {\sqrt{1+2\beta\sigma^{2}D_{0}^{2}}}\cdot \mathcal R_{\rm trunc}(\alpha^*,\beta,\sigma),
  \label{eq:Ssigma_D0_mixed}
\end{align}

where $\alpha^*=\alpha + 2\beta D_0$, and $\mathcal R_{\rm trunc}(\alpha,\beta,\sigma)$ is the same truncation value as derived previously.

Applying this model to the curves yielded  $\mathrm{DEFs_{80\%}}$ of 1.26 (7.7\%) for the 0.25 M concentration, 1.45 (1.2\%) for the 0.50 mM concentration, and 1.68 (-13.4\%) for the 1.00 mM concentration. Percent deviates are in parenthesis.

It's clear that the insertion of the mixed term contributed to an improvement of the $\mathrm{DEFs_{80\%}}$ for both the 0.25 and the 0.50 concentrations, while making it worse for the lowest concentration, meaning that even at low load, the mixed term is not negligible. However, the result for  1.00 mM was still high. 

At higher concentrations, it can be posited that not only the probability of a double baseline/enhanced dose deposition increases, but the probability of two simultaneous cascade-derived dose depositions on the same site also increases, which  this variance-only derivation does not account for.

To account for the two simultaneous hits, the Taylor expansion \ref{eq:Dloc} has to be increased to up to second-order:

\begin{equation}
    D_{\text{enh}} = D_{0}\left(1+\sigma Z + \frac{1}{2}\sigma^2Z^2\right) + \mathcal{O}(\sigma^3).
    \label{eq:second_order}
\end{equation}

To note that this equation can be rewritten as:

\begin{equation}
    D_{0}\left(1+\sigma Z + \frac{1}{2}\sigma^2Z^2\right) = D_0\left(\frac{1}{2}\sigma^2\left(Z+\frac{1}{\sigma} \right)^2+\frac{1}{2}\right),
    \label{eq:second_order_positive}
\end{equation}

which is always positive, therefore there's no need for truncation. Reintroducing this into equation \ref{eq:S_fraction_single}:

\begin{equation}
  \begin{aligned}
-\alpha D_{\text{enh}}
&= -\alpha D_{0}\bigl(1+\sigma Z +\frac{1}{2}\sigma^2 Z^2 \bigr)
   = -\alpha D_{0} \,-\, \alpha\sigma D_{0} Z \,-\, \frac{1}{2}\alpha\sigma D_{0}^2 Z^2, \\[4pt]
-\beta D_{\text{enh}}^{2}
&= -\beta D_{0}^{2}\bigl(1+\sigma Z+\frac{1}{2}\sigma^{2}Z^{2}\bigr)^2  \\[2pt]
&= -\beta D_{0}^{2}\bigl( 1 + 2\,\sigma\,Z + 2\,\sigma^{2}\,Z^{2} + \,\sigma^{3}\,Z^{3} + \frac{1}{2}\sigma^{4}\,Z^{4}\bigr)\\[2pt]
&\approx -\beta D_{0}^{2} \;-\; 2\beta\sigma D_{0}^{2} Z
   \;-\;2\beta\sigma^{2}D_{0}^{2}Z^{2}.
\end{aligned}
\label{eq:coefficients_second_order}
\end{equation}

In which we specifically dropped the terms of higher order than $\mathcal{O}(\sigma^2)$.

Now we can collect the terms of equal powers of $Z$:

\begin{align}
  -\alpha D_{\text{enh}} - \beta D_{\text{enh}}^2
  &= \bigl(-\alpha D_0 - \beta D_0^2 \bigr) +
    \notag\\
  &\quad + \sigma\bigl(-\alpha D_0 - 2\beta D_0^2 \bigr)Z
    + \sigma^2\bigl(-\tfrac12\,\alpha D_0^2 - 2\beta D_0^2 \bigr)Z^2 \,.
    \label{eq:coefficients_second_order_collected}
\end{align}

Defining:
\begin{equation}
\begin{aligned}
a =& -\,\sigma\,D_{0}\,(\alpha + 2\beta\,D_{0})
= -\,\sigma\,D_{0}\,\alpha^{*} \\
b =&  -\,\sigma^{2}\,D_{0}\,\Bigl(\tfrac{1}{2}\,\alpha + 2\beta\,D_{0}\Bigr).
\end{aligned}
\end{equation}

and averaging to obtain the macroscopic survival, yields:

\begin{equation}
    S_{\sigma^2} \;=\; e^{-\alpha D_{0} \;-\; \beta D_{0}^{2}} \; \mathbb{E}\bigl[e^{\,aZ \;+\; bZ^{2}}\bigr], 
    \label{eq:S_sigma_second_order}
\end{equation}

thus being reduced to the integrable Gaussian form. Performing the integration and replacing the values leads to:

\begin{equation}
    S_{\sigma^2} = \frac{\exp\left(-\alpha D_0 - \beta D_0^2 + \frac{\sigma^2 D_0^2(\alpha + 2\beta D_0)^2}{2[1 + \sigma^2 D_0(\alpha + 4\beta D_0)]}\right)}{\sqrt{1 + \sigma^2 D_0(\alpha + 4\beta D_0)}}.
\label{eq:survival_second_order}
\end{equation}

Using this survival curve to determine $\mathrm{DEFs_{80\%}}$ on the three different concentrations, yielding the values of 1.42 (+21.6\%), 1.68 (+14.2\%) and 2.01 (+3.7\%), respectively, with the percent deviations  to experimental values in parentheses.

With this analysis, it becomes clear that the optimal models for each concentration are:

\begin{itemize}
    \item for 0.25 mM and 0.50 mM, the first-order approximation including the mixed term;
    \item for 1.00 mM, the second order approximation
\end{itemize}

\subsection{Model Survival curve comparison for 100 kVp BAEC}

\begin{figure}[h!]
  \centering
  \begin{subcaptionbox}{Results for 0.25 mM and 0.50 using the $S_{\sigma}^{mixed}$ survival curve model.\label{fig:a}}[0.45\linewidth]
    {\includegraphics[width=\linewidth]{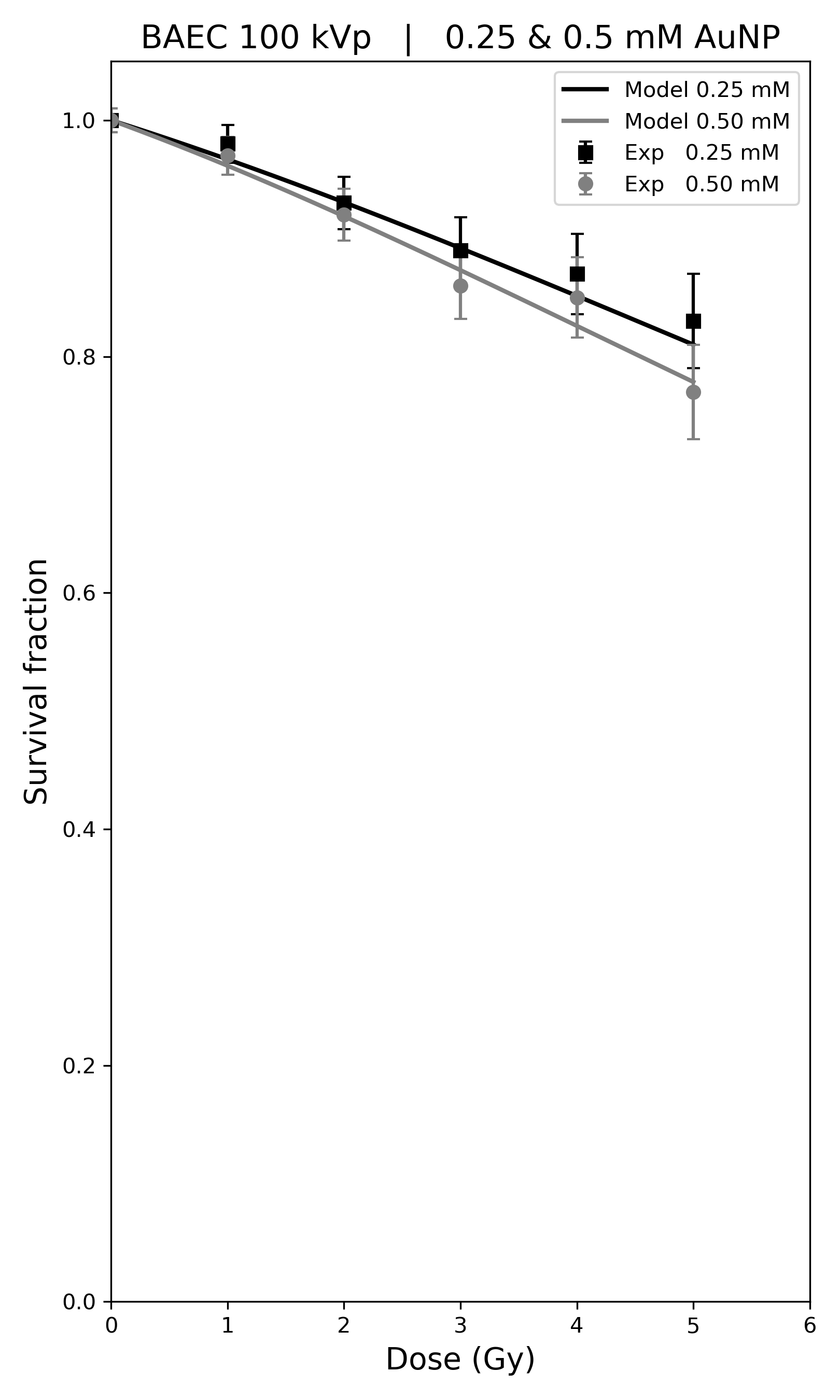}}
  \end{subcaptionbox}
  \hfill
  \begin{subcaptionbox}{Results using the LEM for the same concentrations, taken from \cite{brown2017}.\label{fig:b}}[0.45\linewidth]
    {\includegraphics[width=\linewidth]{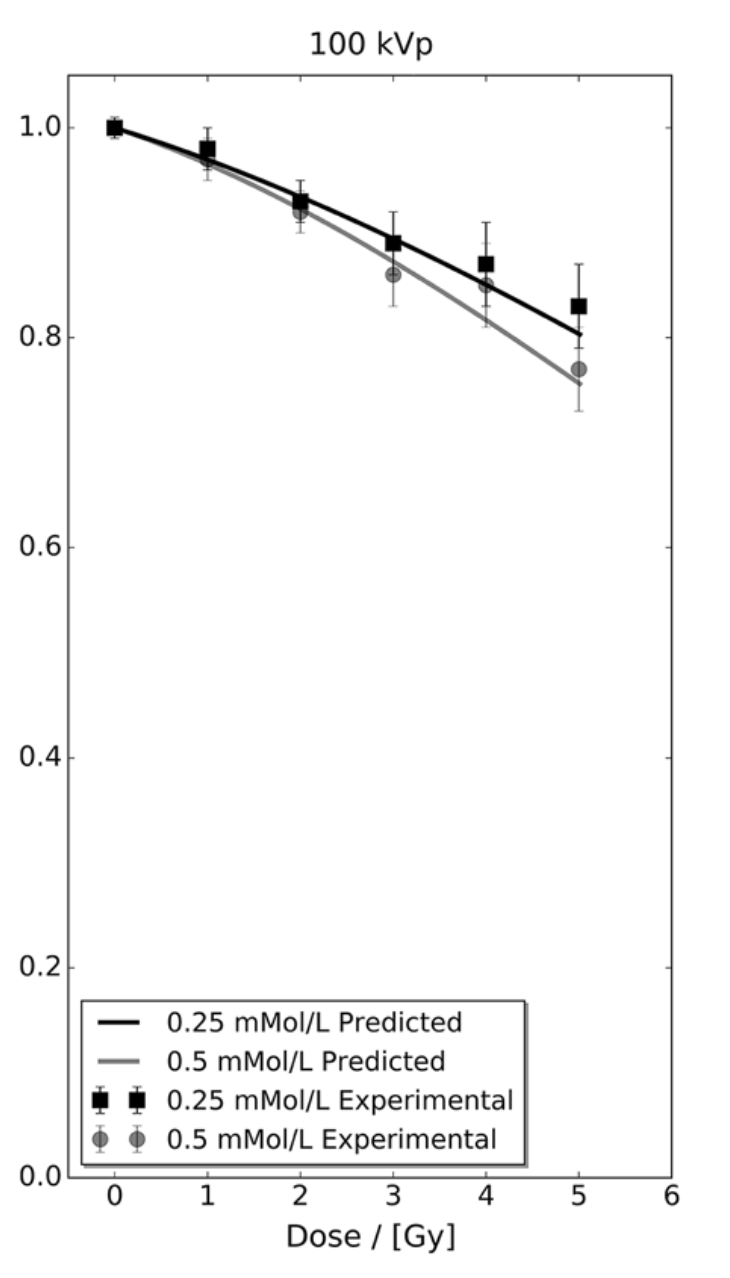}}
  \end{subcaptionbox}
  \caption{Graphical comparison between the analytical models and the LEM \cite{brown2017} for 0.25 and 0.50 mM concentrations}
  \label{fig:quarter_vs_half}
\end{figure}

\begin{figure}[h!]
  \centering
  \begin{subcaptionbox}{Results for 0.00 mM and 1.00 using the $S_{\sigma}$ (for $\sigma=0$ and $S_{\sigma^2}$ survival curve models, respectively.\label{fig:a_II}}[0.45\linewidth]
    {\includegraphics[width=\linewidth]{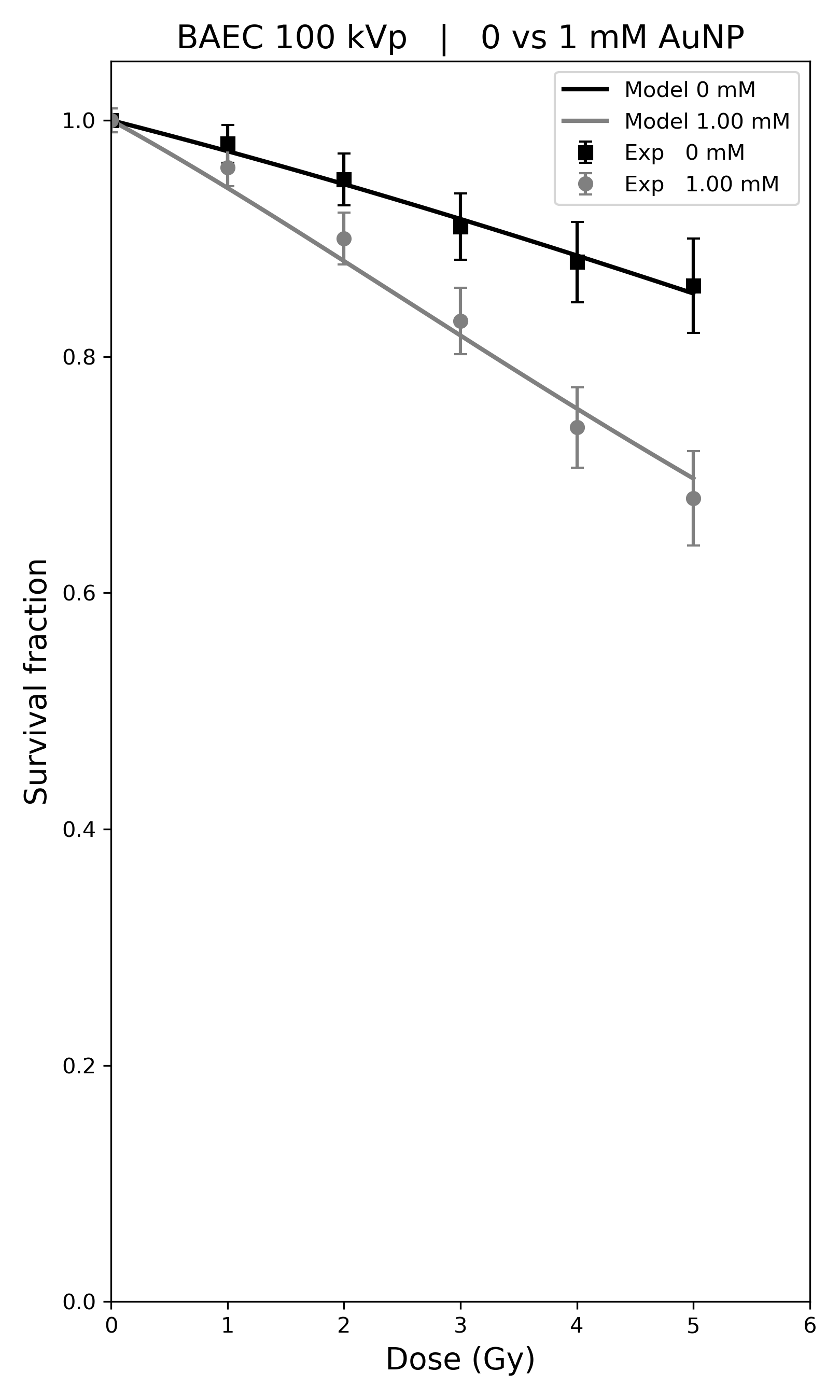}}
  \end{subcaptionbox}
  \hfill
  \begin{subcaptionbox}{Results using the LEM for the same concentrations, taken from \cite{brown2017}.\label{fig:b_II}}[0.45\linewidth]
    {\includegraphics[width=\linewidth]{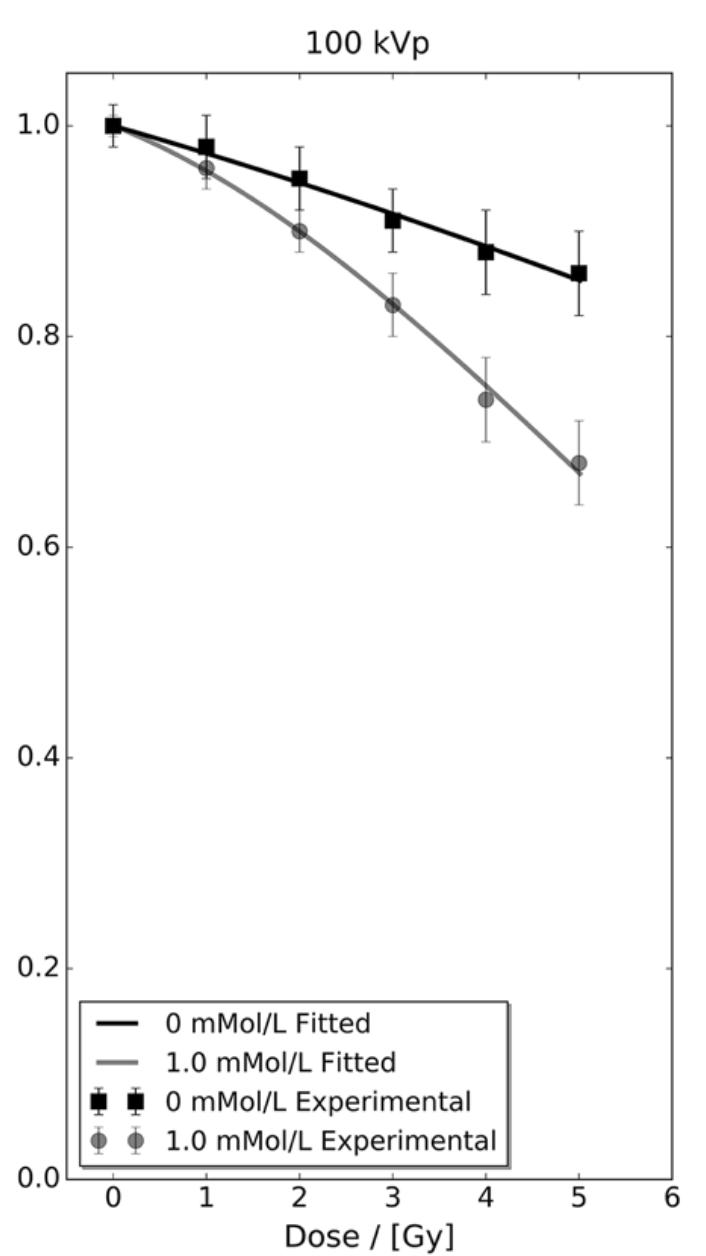}}
  \end{subcaptionbox}
  \caption{Graphical comparison between the analytical models and the LEM \cite{brown2017} for 0.00 and 1.00 mM concentrations}
  \label{fig:zero_vs_one}
\end{figure}

In figure~\ref{fig:zero_vs_one} we present the baseline LQ control fitted with $S_{\sigma}$ for 
(\(\sigma=0\)) and the $S_{\sigma^2}$ model for the 1.00 mM concentration, and in figure~\ref{fig:quarter_vs_half} we compare the $S_{\sigma}^{mixed} $ model curves to the 0.25 mM and 0.50 mM datasets, respectively, side by side with Brown's results for the 100 kVp beam.  

A graphical analysis shows that the obtained survival curves explain the experimental points within a small uncertainty, confirmed by the experimental survival fraction values obtained from \cite{rahman2009, brown2017} compared with the results using our models, shown in table \ref{tab:SF_model_vs_exp}.

It is noticeable that all results fall below percent deviations smaller than 2.5\%. 

Despite the good results of the model, figures \ref{fig:quarter_vs_half} and \ref{fig:zero_vs_one} reveal that the obtained survival curves show less curvature when compared to the classical LEM curves.

\begin{landscape}
\begin{table}[h!]
\centering
\begin{tabular}{c|ccc|ccc|ccc|ccc}
\toprule
\multirow{2}{*}{Dose (Gy)} 
& \multicolumn{3}{c|}{0 mM} 
& \multicolumn{3}{c|}{0.25 mM} 
& \multicolumn{3}{c|}{0.50 mM} 
& \multicolumn{3}{c}{1.00 mM} \\
& SF\textsubscript{exp} & SF\textsubscript{model} & \% dev
& SF\textsubscript{exp} & SF\textsubscript{model} & \% dev
& SF\textsubscript{exp} & SF\textsubscript{model} & \% dev
& SF\textsubscript{exp} & SF\textsubscript{model} & \% dev \\
\midrule
0.0 & 1.00 & 1.0000 &  0.0 & 1.00 & 1.0000 &  0.0 & 1.00 & 1.0000 &  0.0 & 1.00 & 1.0000 &  0.0 \\
1.0 & 0.98 & 0.9738 & -0.6 & 0.98 & 0.9669 & -1.3 & 0.97 & 0.9615 & -0.9 & 0.96 & 0.9427 & -1.8 \\
2.0 & 0.95 & 0.9459 & -0.4 & 0.93 & 0.9306 &  0.1 & 0.92 & 0.9187 & -0.1 & 0.90 & 0.8808 & -2.1 \\
3.0 & 0.91 & 0.9164 &  0.7 & 0.89 & 0.8917 &  0.2 & 0.86 & 0.8731 &  1.5 & 0.83 & 0.8178 & -1.5 \\
4.0 & 0.88 & 0.8855 &  0.6 & 0.87 & 0.8513 & -2.1 & 0.85 & 0.8259 & -2.8 & 0.74 & 0.7559 &  2.2 \\
5.0 & 0.86 & 0.8534 & -0.8 & 0.83 & 0.8101 & -2.4 & 0.77 & 0.7785 &  1.1 & 0.68 & 0.6968 &  2.5 \\
\bottomrule
\end{tabular}

\caption{Experimental and model survival fractions (SF) and percentage differences for each AuNP concentration.}
\label{tab:SF_model_vs_exp}
\end{table}
\end{landscape}

Table~\ref{tab:lqfits} lists the
\(\alpha',\beta'\) pairs obtained by fitting the analytic survival
curves with a linear–quadratic law
\(\ln S = -\alpha' D-\beta' D^{2}\) using a least-squares fit in linear space.
Errors are $1\,\sigma$ from the covariance matrix; \(R^{2}\geq0.999\) in
all cases.

\begin{table}[h!]
\centering
\begin{tabular}{@{}cccccc@{}}
\toprule
Conc.\,(mM) & $\alpha'\;(\mathrm{Gy}^{-1})$
            & $\beta'\;(\mathrm{Gy}^{-2})$
            & $R^{2}$ &
            \multicolumn{2}{c}{Brown\,(100\,kVp)} \\[2pt]
            & $\pm$SE & $\pm$SE & & $\alpha'$ & $\beta'$ \\ \midrule
0.00 & 0.0252\,$\pm$\,0.0000 & 0.00130\,$\pm$\,0.00000 & 1.000 & – & – \\
0.25 & 0.0323\,$\pm$\,0.0001 & 0.00195\,$\pm$\,0.00002 & 1.000 & – & – \\
0.50 & 0.0380\,$\pm$\,0.0002 & 0.00240\,$\pm$\,0.00004 & 0.999 & – & – \\
1.00 & 0.0590\,$\pm$\,0.0004 & 0.00251\,$\pm$\,0.00009 & 0.999 & 0.0347 & 0.00904 \\ \bottomrule
\end{tabular}
\caption{LQ parameters fitted to the $\sigma$-LEM curves.
         The last column shows Brown’s empirical
         \(1.00\,\mathrm{mM}\) pair for comparison.}
\label{tab:lqfits}
\end{table}

At 1.00 mM the analytic curve yields \((\alpha',\beta')=(0.0590,\,0.00251)\), which are completely different from the values of (0.0347, 0.00904), from \cite{brown2017}. This difference will be exploited further in the discussion section.

\section{Discussion}

During derivation, we treated 
$\varepsilon_{\mathrm{cas}}N_{\mathrm{Auger}}$ as an
energy-independent constant.  This is a first-order
simplification: the number of primary photo-ionisations in a nanoparticle
is likely Poisson-distributed and therefore energy-dependent, so future work should focus on a systematic,
energy-resolved derivation, ideally guided by Monte-Carlo results.

A mid-step in the derivation of the survival curves was the calculation of the beam quality constant $K_{c}$ which, due to the value $\varepsilon_{\mathrm{cas}}N_{\mathrm{Auger}}$ being unknown to good accuracy, was calibrated against synchrotron monoenergetic beam results. Results showed that $K_{c}$ peaks in the kV range ($3.22 \,\text{mM}^{-1}$ at 50\,kVp, $2.11\,\text{mM}^{-1}$ at 100\,kVp) and falls to $\mathcal{O}(10^{-4})\,\text{mM}^{-1}$ for MV spectra,  faithfully reproducing the measured dip at 70 keV, related with gold's L3 / K-edges. The unrealistic DER predictions at the MV regime point to the current formalism's limitation to \(\lesssim150-200\) keV.

The derivation of the survival curves relied on assuming a log-normal distribution for the dose enhancement, driven solely by stochasticity. Although a closed form is not attainable using the log-normal distribution, Taylor expansion to first and second order, with or without truncation, allowed for the development of three variants of the model, the  $S_{\sigma}$, which is adapted for low concentration regimes, with sparse local dose enhancements, dominated by single Auger cascade hits and low probability of a simultaneous baseline and Auger cascade hit happening on the same site as being negligible. The $S_{\sigma}^{mixed}$ model is well adapted to an intermediate concentration level, which, although still dominated by single Auger cascade hits, the mixed simultaneous hit term  can no longer be neglected. Finally, the $S_{\sigma^2}$ model is adapted to higher concentrations. Here, the Taylor expansion is taken to second order, therefore allowing both single and double Auger cascade hits to dominate the dose enhancement.

The models were compared with previous results obtained for BAEC cells irradiated with 100 kVp, both with experimental and LEM derived data\cite{brown2017}. Even despite the fact that the X-ray spectrum used for the BAEC results was different from the spectrum we used for our model, the value of $K_c=2.11$ mM$^{-1}$ was well adapted to the results. 

The models were able to reproduce DEF$_{80\%}$ to good accuracy, although less so for the lower concentration  0.25 mM dataset, because the averaging over the heterogeneity of dose concentrations is less valid for lower concentrations (where the probability of "no-gold" uptake in cells is more likely), making the $\sigma$-LEM model overestimate the value of enhancement for these cases. In any case, the mixed curves optimise the 0.25, 0.50 mM datasets, and the second--order is well adapted to the 1.00\,mM dataset. 

Fitted LQ parameters show that enhancement is predominantly $\alpha$--driven and less curved than standard LEM predictions, at least for the studied case, unlike what is assumed in\cite{brown2017}. 

In fact, in our model, we assume that $\varepsilon_{\mathrm{cas}}N_{\mathrm{Auger}}$ leads to single Auger cascades producing single--track events, with or without a mixed term which also deals with single-track hits probability of being simultaneous with a baseline hit. The mixed term introduction lowers $\beta'$ as it literally implies a growth in $\alpha$ by $2\beta D_0$. Two cascades must overlap to improve upon $\beta$, which is impossible in the $S_{\sigma}$ and the $S_{\sigma}^{mixed}$ models. Only in the second-order expansion are double cascades accounted for.

Thus the models in these concentration ranges and beam quality seem to indicate radiosensitisation is chiefly $\alpha$--driven, with $\beta$ playing only a secondary role at the only concentration deemed high (1.00 mM) \cite{McMahon2011,Subiel2016}. This is in contrast with the LEM approach used in \cite{brown2017}, which is in fact blind to this issue, as it scores the dose depositions regardless of their origin and then averages over the entire domain. The $\sigma$-LEM model exposes exactly where $\beta$ dominated events become relevant (at second-order, $\sigma^2$ terms), pointing to a physical interpretation of when the linear or the quadratic term should dominate. The sparser Auger cascade-driven enhancement has a lower probability of two $\beta$-style double-hits, which only become relevant at higher concentrations. 

Using Brown's phenomenological derivation:
\begin{equation*}
    \alpha(C)=\alpha_{U}+\frac{C}{C_{0}}\Delta\alpha,\qquad
\beta(C)=\beta_{U}+\frac{C}{C_{0}}\Delta\beta.
\end{equation*}

Keeping only the $\alpha$ term for simplicity, the survival fraction is  

\begin{equation*}
    \mathrm{SF}(C,D)=
\exp\!\bigl[-\alpha(C)D\bigr]
=\exp\!\bigl[-\alpha_{U}D\bigr]\,
\exp\!\Bigl[-\tfrac{\Delta\alpha}{C_{0}}\,C\,D\Bigr].
\end{equation*}

For small concentrations $C/C_{0}$ the second exponential can be
linearised:

\begin{equation*}
    \exp\!\Bigl[-\tfrac{\Delta\alpha}{C_{0}}\,C\,D\Bigr]
\simeq
1-\frac{\Delta\alpha}{C_{0}}\,C\,D,
\end{equation*}

giving
\begin{equation*}
    \mathrm{SF}(C,D)\simeq
e^{-\alpha_{U}D}\!
\Bigl[1-\tfrac{\Delta\alpha}{C_{0}}\,C\,D\Bigr].
\end{equation*}

If we consider a  reference survival level $\mathrm{SF}^{\star}$  in the absence of AuNPs
at dose $D_{0}$
\(\mathrm{SF}^{\star}=e^{-\alpha_{U}D_{0}}\).

The same survival can be obtained at dose $D_{C}$ in the presence of a concentration of $C$ AuNPs , so we can write:

\begin{equation*}
    e^{-\alpha_{U}D_{0}}=
e^{-\alpha_{U}D_{C}}
\Bigl[1-\frac{\Delta\alpha}{C_{0}}\,C\,D_{C}\Bigr].
\end{equation*}

Considering $D_C=D_0+\delta D$, and dividing both sides by $e^{\alpha_U(D_0+\delta D)}$ yields:

\begin{equation*}
    e^{\alpha_U\delta D}=1-\frac{\Delta\alpha}{C_{0}}\,C\,(D_{0}+\delta D)
\end{equation*}

Expanding the left side to first order:

\begin{equation*}
    1-\alpha_U\delta D=1-\frac{\Delta\alpha}{C_{0}}D_0-\frac{\Delta\alpha}{C_{0}}\delta D,
\end{equation*}

and neglecting the term in $\delta D$, one obtains
\[
1-\alpha_{U}\delta D+\tfrac{\Delta\alpha}{C_{0}}\,C\,D_{0}=1
\quad\Longrightarrow\quad
\delta D=\frac{\Delta\alpha}{\alpha_{U}}\frac{C}{C_{0}}\,D_{0}.
\]
Finally,
\[
\;
\mathrm{DER}\equiv\frac{D_{C}}{D_{0}}
=\frac{D_{0}+\delta D}{D_{0}}
=1+\frac{\delta D}{D}= 1+\frac{\Delta\alpha}{\alpha_{U}}\frac{C}{C_{0}}.
\tag{A4}
\]

which is algebraically identical to equation \ref{eq:DERc}, in which $K_c=\frac{1}{C_0}\frac{\Delta\alpha}{\alpha_{U}}$.

A systematic survey of current literature confirmed that no earlier work combined a log-normal description of dose with a variance-driven Local-Effect-Model ($\sigma$-LEM).  Prior LEM-based approaches either relied on full Monte-Carlo track scoring \cite{McMahon2011,lechtman2013} or on phenomenological interpolations that tuned the LQ coefficients directly to concentration \cite{brown2017}. 

As mentioned previously, recent research \cite{Blind2025} assumes a similar relation between enhanced dose and concentration (equation 1 in their work) in nanoparticle enhancement, although this is used as a tool to improve Monte Carlo or survival curves by adding an extra phenomenological exponential term to the survival curve.

Also, Pan \textit{et al}\cite{Pan2022} used a complex algebraic derivation from compound Poisson distributions to recover first-order linear increases in the $\alpha$ term, yet does not invoke Gaussian integrals or moment expansions to obtain closed-form dose-enhancement ratios.

The absence of a previous variance-based formulation reinforces the originality of our analytic derivation, while the qualitative agreement with Brown’s interpolation lends external support to the key prediction that \(\mathrm{DER}=1+K_c\,c\) at low to moderate concentrations. In addition, the finding that radiosensitisation is predominantly $\alpha$-driven is compatible with Monte-Carlo simulation results \cite{lechtman2013}.

\section{Conclusions}

The analytical $\sigma$-LEM was successful in modeling the BAEC experimental results, with $DEF_{80\%}$ below $5\%$ for all concentration regimes, and reproducing survival curves with percent deviations to the experimental values smaller than $2\%$, except for one point.

Closed-form expressions for the survival curves were attainable by expanding to first- and second-order moments of the log-normal distribution. The three variants of the model, for low, intermediate, and high concentrations, reveal that sensitisation is mainly $\alpha$-driven, at least until the high concentration regime.

The great advantage of this model is that it collapses radiosensitisation to a beam-quality related factor,\(K_{c}(E)\), cell geometry (or at least size of nucleus) and intracellular concentration \(c\), inferring the survival curves from first principles, allowing for better insights relative to the physical mechanisms behind dose enhancement.

Uncertainty in the $K_{c}(E)$ value entailed the need to calibrate it to synchrotron monoenergetic data. 

Also, the fact that the model predicts DERs close to 1 in the MV range indicates a limit of validity of the model.

Future work should refine the treatment of $\varepsilon_{\mathrm{cas}}N_{\mathrm{Auger}}$, attempt to model the MV regimes, but most importantly, test the formalism for other beam qualities, cell lines, and in-vitro/in-silico scenarios.

\bibliographystyle{elsarticle-num}
\bibliography{references}

\clearpage
\section*{Supplementary Material}
\appendix

\section{Expansion of $\mathbb E_{Z}\!\bigl[
        e^{-\alpha\sigma D_{0}Z-\beta\sigma^{2}D_{0}^{2}Z^{2}}
     \bigr]$}

In this section we will show that the following identity holds:
\[
\mathbb{E}_Z \left[ e^{tZ + bZ^2} \right] = \frac{1}{\sqrt{1 - 2b}} \exp\left( \frac{t^2}{2(1 - 2b)} \right), \quad b < \frac{1}{2},
\]

with $t=-\alpha\, \sigma\, D_0$ and $b=-\beta \,\sigma^2\,D_0^2$.

The expectation value of a given function $f(Z)$ in which \( Z \sim \mathcal{N}(0,1) \) can be written as:

\[
\mathbb{E}_{f(Z)} =  \frac{1}{\sqrt{2\pi}} \int_{-\infty}^{\infty} f(z)\,e^{-\frac{z^2}{2}} \, dz.
\]

Replacing
\[
\mathbb{E} \left[ e^{tZ + bZ^2} \right] = \frac{1}{\sqrt{2\pi}} \int_{-\infty}^{\infty} e^{t z + b z^2 - \frac{z^2}{2}} \, dz
= \frac{1}{\sqrt{2\pi}} \int_{-\infty}^{\infty} e^{-(a z^2 + b' z + c)} \, dz,
\]

in which the right-hand side of the equation is equation 7.4.32 from \cite{abramowitz1965handbook},

Completing the square leaves us with:

\[tz+bz^2-\frac{z^2}{2}=-(\frac{1}{2}-b)z^2+tz=-az^2+tz.
\]

\begin{align*}
    -az^{2}+tz & =\\
    & = -a\!\Bigl[z^{2}-\frac{t}{a}\,z\Bigr] \\
    & = -a\!\Bigl[\Bigl(z-\frac{t}{2a}\Bigr)^{2}-\frac{t^{2}}{4a^{2}}\Bigr] \\
    & = -a\Bigl(z-\frac{t}{2a}\Bigr)^{2}+\frac{t^{2}}{4a} \\
    & =-a\Bigl[\Bigl(z-\frac{t}{2a}\Bigr)^{2}-\frac{t^{2}}{4a}\Bigr].
\end{align*}

Identifying the terms with equation 7.4.32, \( a = \frac{1}{2} - b \), \( b' = 0 \), and \(c = -\frac{t^2}{2(1 - 2b)}\), leaves us with:

\[\mathbb{E} \left[ e^{tZ + bZ^2} \right]   =\frac{1}{\sqrt{2\pi}} \int_{-\infty}^{\infty} e^{-(a z'^2 + c)} \, dz',
\]

where we have made the change of variable $z'=z-\frac{t}{2a}$.
The value of $c$ is constant in \(z'\), and can therefore be taken out of the integral, leaving:

\[
\mathbb{E} \left[ e^{tZ + bZ^2} \right] = \frac{1}{\sqrt{2\pi}}\,e^{-c} \int_{-\infty}^{\infty} e^{-a z'^2} \, dz'.
\]

Now replacing in equation 7.4.32:
\[
\int e^{-a z'^2 } \, dz'= \frac{1}{2}\sqrt{\frac{\pi}{a}} \, \text{erf}\left( \sqrt{a} z'  \right) + \text{const.}
\]

For the definite integral over \( ]-\infty, \infty[ \), the error function term simplifies, because when $z'\rightarrow+\infty$, $\text{erf}(f(z'))\rightarrow1,$ and vice-versa, $z'\rightarrow-\infty$, $\text{erf}(f(z'))\rightarrow-1$. We then obtain:

\[
\frac{e^{-c}}{\sqrt{2\pi}}\int_{-\infty}^{\infty} e^{-a z'^2} \, dz' = \frac{e^{-c}}{\sqrt{2\pi}}\sqrt{\frac{\pi}{a}}, \quad a > 0
\]

replacing $c=-\frac{t^2}{2(1-2b)}$, and $a=\frac{1}{2}-b$, therefore retrieving $b$:

\[
\mathbb{E} \left[ e^{tZ + bZ^2} \right] =\frac{1}{\sqrt{2\pi}}\,\sqrt{\frac{\pi}{\frac{1}{2}-b}}\,\exp\left( \frac{t^2}{2(1-2b)} \right)
\]

Simplifying and thus confirming:
\[
\mathbb{E} \left[ e^{tZ + bZ^2} \right] = \frac{1}{\sqrt{1 - 2b}} \exp\left( \frac{t^2}{2(1 - 2b)} \right)
\]

with $b < \frac{1}{2}$ because $a>0$.

The integral for the second-order moment can be derived in a similar fashion.

\section{Derivation of the correction factor for the truncated gaussian}

For the first–order dose we keep only the Gaussian tail  
\(Z\ge-1/\sigma\).  
Let  

\[
P(\sigma)=\Pr[Z\ge -1/\sigma]=\tfrac12\,\operatorname{erfc}\!\bigl(L/\sqrt2\bigr)
\]

so that, instead of integration over the full domain $]-\infty,+\infty[$, we integrate over   $]-1/\sigma,+\infty[$. For that we need to scale by the factor $P(\sigma$):

\[
\frac{1}{P(\sigma)}\,
\frac{1}{\sqrt{2\pi}}\int_{1/\sigma}^{\infty}
      e^{t z + b z^{2}-z^{2}/2}\,\mathrm dz ,
\qquad b<\tfrac12 .
\]

Using the
standard upper-limit Gaussian integral on equation \ref{eq:D_enh_first_order} gives  

\[
      e^{-\alpha D_{0}-\beta D_{0}^{2}}\,
     \frac{
           \exp\Bigl(
             \dfrac{\alpha^{2}\sigma^{2}D_{0}^{2}}
                   {2\bigl(1+2\beta\sigma^{2}D_{0}^{2}\bigr)}
           \Bigr)}
          {\sqrt{1+2\beta\sigma^{2}D_{0}^{2}}}\mathcal\cdot R_{\rm trunc}(\alpha,\beta,\sigma),
\]

where the correction factor $\mathcal R_{\rm trunc}(\alpha,\beta,\sigma)$ is given by:

\[
\mathcal R_{\text{trunc}}(\alpha,\beta,\sigma)
  \;=\;
  \frac{
    \operatorname{erfc}\!\biggl(
      \frac{\sqrt{1+2\beta\sigma^{2}D_{0}^{2}}}{\sqrt{2}}\,
      \biggl[
        -\frac{1}{\sigma}
        +\frac{\sigma D_{0}\alpha}
               {1+2\beta\sigma^{2}D_{0}^{2}}
      \biggr]
    \biggr)}
    {\operatorname{erfc}\!\bigl(-\tfrac{1}{\sigma\sqrt{2}}\bigr)}
\]

\section{Breakdown of the Dose Enhancement Factor at 80\% survival results}

\begin{table}[h!]
\centering
\caption{DEF$_{80}$ comparison across $\sigma$-LEM model hierarchy.; baseline D$_{80}$ = 6.605\,Gy.}
\label{tab:def80}
\begin{tabular}{@{}c|c|ccc@{}}\toprule
\multirow{2}{*}{Conc. (mM)} & \multirow{2}{*}{Exp.} & \multicolumn{3}{c}{$\sigma$-LEM models}\\
\cmidrule(l){3-5}
& & Variance & Mixed & Complete 2nd\\\midrule
0.25 & 1.170 & 1.301 & 1.260 & 1.422\\
0.50 & 1.470 & 1.504 & 1.453 & 1.679\\
1.00 & 1.940 & 1.738 & 1.680 & 2.011\\
\bottomrule
\end{tabular}
\end{table}

\begin{table}[h!]
\centering
\caption{Percentage errors with respect to experimental DEF$_{80}$.}
\label{tab:def80err}
\begin{tabular}{@{}c|ccc@{}}\toprule
Concentration (mM) & Variance & Variance + Mixed & Complete 2nd\\\midrule
0.25 & +11.2 & +7.7 & +21.6\\
0.50 & +2.3 & -1.2 & +14.2\\
1.00 & -10.4 & -13.4 & +3.7\\
\bottomrule
\end{tabular}
\end{table}

\begin{table}[h!]
\centering
\caption{Best model (minimum |error|) per concentration.}
\label{tab:bestmodel}
\begin{tabular}{@{}c|c|c@{}}\toprule
Concentration (mM) & Best model & Error (\%)\\\midrule
0.25 & Variance + Mixed term & +7.7\\
0.50 & Variance + Mixed term & +1.2\\
1.00 & Complete 2nd & +3.7\\
\bottomrule
\end{tabular}
\end{table}

\end{document}